\documentclass[
aps,
apr,
amsmath,amssymb,
preprint,
superscriptaddress,
amsart,
]{revtex4-2}

\usepackage{graphicx}
\usepackage[utf8]{inputenc}
\graphicspath{ {images/} }
\usepackage{units}
\usepackage{xcolor}
\usepackage{tabularx}
\usepackage[percent]{overpic}
\usepackage{bm}

\usepackage{siunitx}
\usepackage{multirow}
\usepackage{verbatim}

\begin{document}

\title{
Microacoustic metagratings at ultra-high frequencies fabricated by two-photon lithography}

\author{Anton Melnikov}
\email{anton.melnikov@ipms.fraunhofer.de}
\affiliation{
Fraunhofer Institute for Photonic Microsystems IPMS, Dresden, Germany}

\author{Sören Köble}
\affiliation{
Fraunhofer Institute for Photonic Microsystems IPMS, Dresden, Germany}

\author{Severin Schweiger}
\affiliation{
Fraunhofer Institute for Photonic Microsystems IPMS, Dresden, Germany}

\author{Yan Kei Chiang}
\affiliation{
School of Engineering and Information Technology\\
University of New South Wales, Canberra, Australia}

\author{Steffen Marburg}
\affiliation{
Chair of Vibro-Acoustics of Vehicles and Machines\\
Technical University of Munich, Germany}

\author{David A.~Powell}
\affiliation{
School of Engineering and Information Technology\\
University of New South Wales, Canberra, Australia}

\begin{abstract}

The recently proposed bianisotropic acoustic metagratings offer promising opportunities for passive acoustic wavefront manipulation, which is of particular interest in flat acoustic lenses and ultrasound imaging at ultra-high frequency ultrasound.
Despite this fact, acoustic metagratings have never been scaled to MHz frequencies that are common in ultrasound imaging. One of the greatest challenges is the production of complex structures of microscopic size.
Owing to two-photon polymerization, a novel fabrication technique from the view of acoustic metamaterials, it is now possible to precisely manufacture sub-wavelength structures in this frequency range.
However, shrinking in size poses another challenge; the increasing thermoviscous effects lead to considerable losses, which must be taken into account in the design.
In this work we propose three microacoustic metagrating designs refracting a normally incident wave towards $\unit[-35]{^{\circ}}$ at \unit[2]{MHz}.
In order to develop metaatoms insensitive to thermoviscous effects we use shape optimization techniques incorporating the linearized Navier-Stokes equations discretized with finite element method.
We report for the first time microscopic acoustic metamaterials manufactured using two-photon polymerization and, subsequently, experimentally verify their effectively using a capacitive micromachined ultrasonic transducer as source and an optical microphone as a detector in a range from \unit[1.8]{MHz} to \unit[2.2]{MHz}.
We demonstrate not just that a microacoustic metagrating can effectively redirect the normally incident wave despite the thermoviscous losses, but also that it being only $0.29\lambda$ thick can allocate \unit[90]{\%} of the transmitted energy in the $-1$st diffraction order.
\end{abstract}

\maketitle

\section{Introduction}

Acoustic metamaterials make it possible to design unique material properties that do not occur in nature \cite{Cummer.2016, Haberman.2016, Ma.2016, Jimenez.2021}, for example negative bulk modulus and negative dynamic mass density \cite{Fang.2006, Koo.2016}.
Because of these unusual properties, acoustic metamaterials have drawn attention in the context of unconventional acoustic wave manipulation.
This opened up a wide range of applications, such as acoustic cloaking \cite{Farhat.2009, Sanchis.2013, Zigoneanu.2014}, acoustic barriers of subwavelength thickness \cite{Cheng.2015, Wu.2018, Melnikov.2020}, flat acoustic lenses \cite{Kaina.2015, Chen.2018, Jin.2019}, and ultrasonic imaging with subwavelength resolution \cite{Li.2009, Zhu.2011, Cheng.2013}.
In the case of the latter two examples, the class of gradient metasurfaces including reflecting \cite{Li.2014, Guo.2021} and refracting \cite{Xie.2014, Moleron.2014, Liu.2017, Lan.2017, Lan.2017b, Li.2018b, Zhang.2020} designs were proven by experiments to be an effective passive approach for wavefront control, e.g. for acoustic Fresnel lenses \cite{Moleron.2014, Li.2014, Lan.2017, Lan.2017b, Zhang.2020}.
However, the wavelength limits the resolution and therefore for high resolution applications the frequency rises to high or even ultra-high frequency ultrasound requiring scaling of the metasurface geometries.
When speaking of airborne ultrasound, the ultra-high frequency range starts from \unit[0.5]{MHz}, where thermoviscous layers at sound-hard boundaries introduce stronger losses \cite{Ward.2015, Cotterill.2018}.
Because gradient acoustic metasurfaces use very thin acoustic channels and are often enhanced by Helmholtz resonators, the thermoviscous effects can become severe enough to interfere with the proper functioning of the design.
Bianisotropic metagratings \cite{Torrent.2018, Ni.2019, Hou.2019, Craig.2019, Chiang.2020, Chiang.2021} are promising for this ultra-high frequency range, since their geometrical features are close to the wavelength and, therefore, much larger than in the case of gradient metasurfaces.
However, such scaling has never been investigated due to challenges in manufacturing such microscopic structures and the need to make the design insensitive to thermoviscous losses.
Due to the latest developments in the additive manufacturing of microscopic structures, in particular two-photon lithography, one of the biggest hurdles on the way to the experimental realization of microacoustic metagratings can now be overcome.

\begin{figure}[hbtp]
	\textbf{a} \includegraphics[height = 5.0cm]{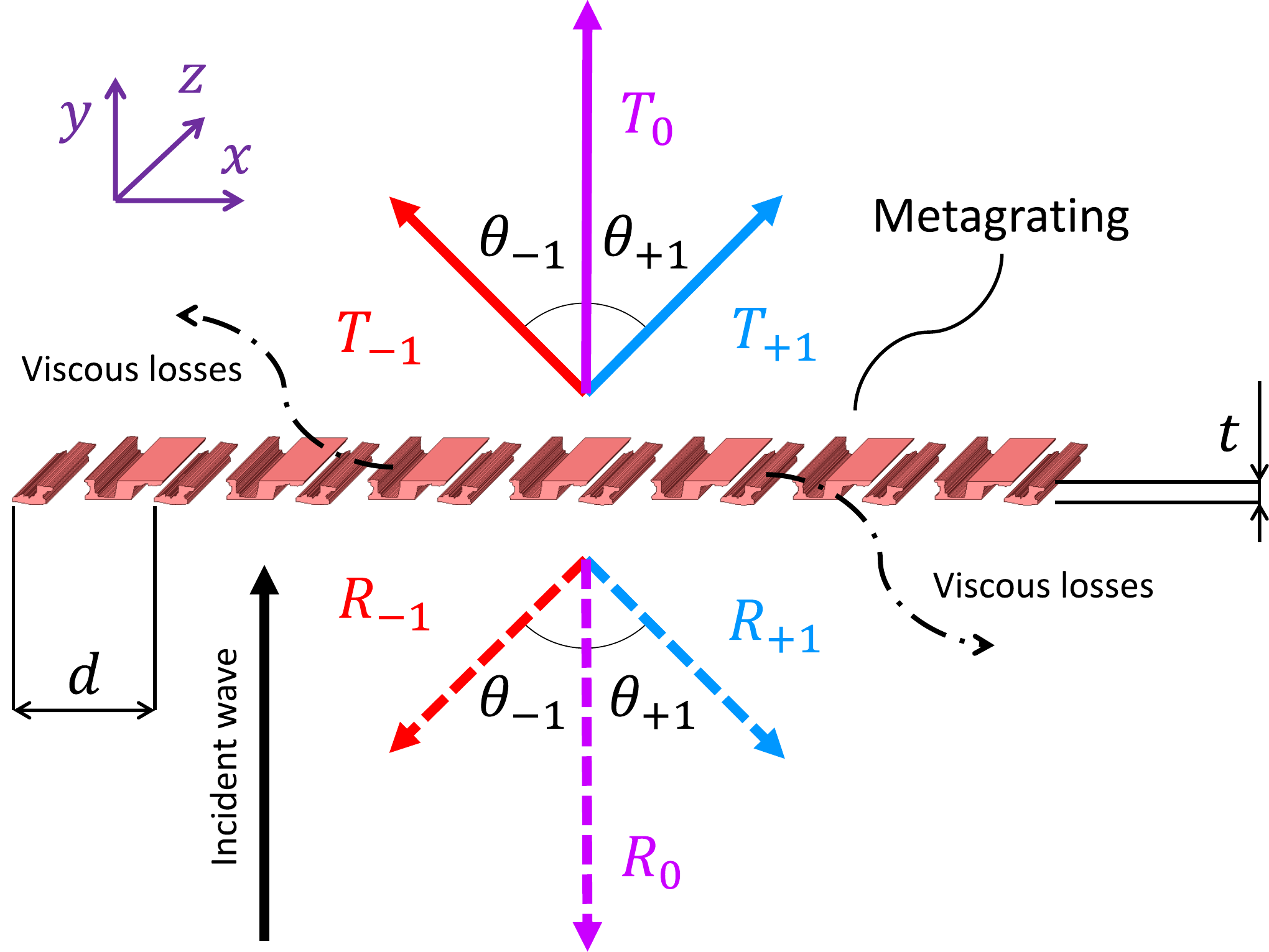}\\
	\textbf{b} \includegraphics[height = 5.0cm]{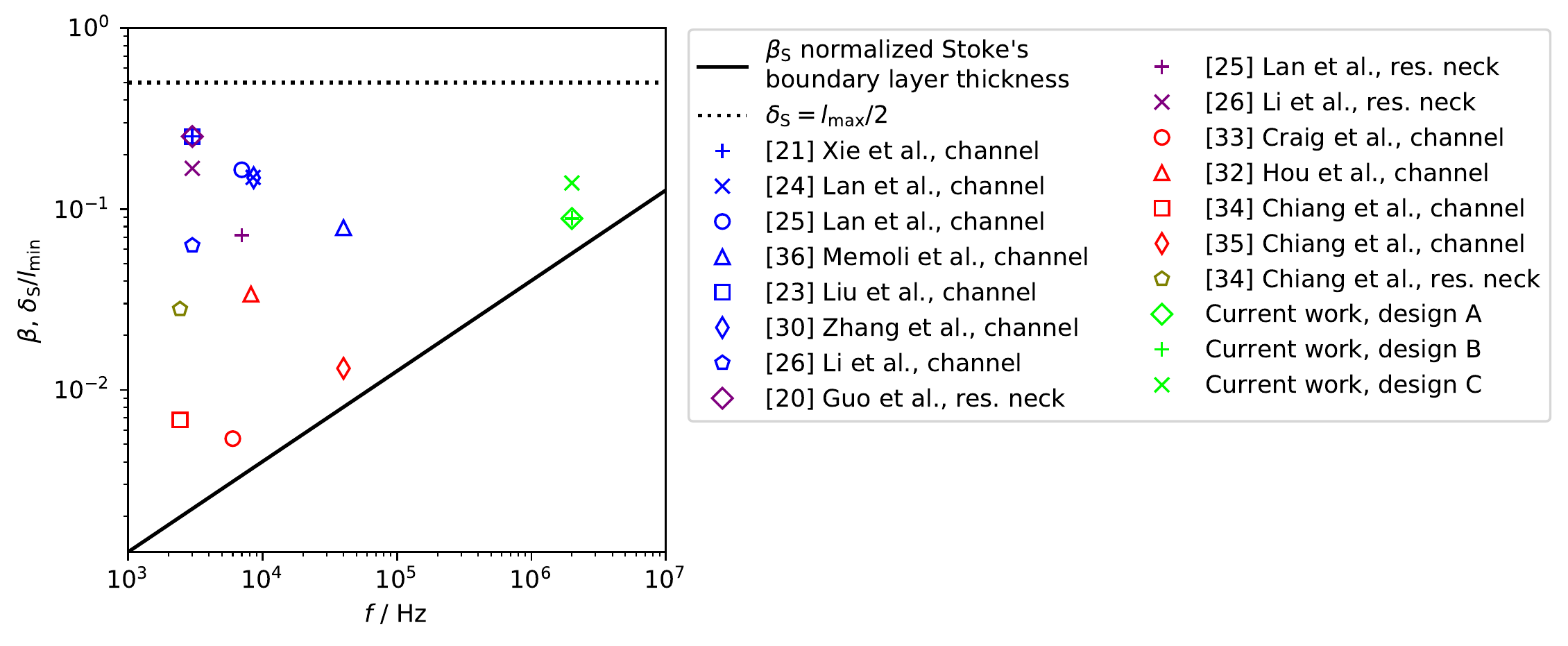}
	\caption{
	    \textbf{Wavefront directions generated by microacoustic metagrating and design aspects.}
	    \textbf{a}
	    Microacoustic metagrating with three diffraction orders and viscous losses with a normal incident wave.
	    Refracted and reflected wavefronts are indicated by solid and dashed arrows, respectively.
	    The diffraction order is indicated by the subscripts.
	    The global metagrating parameters are the lattice constant $d$ and the grating thickness $t$.
	    \textbf{b}
	    Normalized Stoke's boundary layer thickness $\beta_{\mathrm{S}}$
	    and Stoke's boundary layer $\delta_{\mathrm{S}}$ compared to the thinnest channel dimension $l_{\mathrm{min}}$ for a selection of acoustic metamaterial structures taken from the literature
	    \cite{Xie.2014, Lan.2017, Lan.2017b, Memoli.2017, Liu.2017, Zhang.2020, Li.2018b, Guo.2021, Craig.2019, Chiang.2020, Chiang.2021}.
	    \label{fig:meta_basic}}
\end{figure}

Acoustic Metagratings \cite{Radi.2021} are periodic arrays of discrete metaatoms, which can diffract the incident acoustic energy via multiple diffraction orders as shown in Fig.~\ref{fig:meta_basic}a. The number of modes and diffraction angles are determined by the acoustic wavelength $\lambda$ and the metagrating lattice constant $d$, based on Bragg's condition, i.e., $d = n \lambda / \sin{\theta_{n}}$.
The energy redirection to different diffraction orders is attributed to the scattering properties of the meta-atoms and is determined by their geometric shape.
It has recently been shown that by including an additional degree of freedom, represented by the Willis coupling parameter, the acoustic analogy of bianisotropy in electrodynamics, more efficient metamaterial designs can be created \cite{Willis.2011, Radi.2017, Muhlestein.2017, Quan.2018, Melnikov.2019}.
A non-zero Willis coupling parameter is linked to geometric asymmetry of the metaatom \cite{Quan.2018, Melnikov.2019} and in case of metagratings it decouples the $+n$th and $-n$th diffraction orders from each other.
Based on this mechanism, acoustic metagratings have been experimentally demonstrated in the audible frequency range; at \unit[6]{kHz} by Craig \textit{et al.}\cite{Craig.2019}, at \unit[8.2]{kHz} by Hou \textit{et al.}\cite{Hou.2019}, and in the range from \unit[2.43]{kHz} to \unit[2.54]{kHz} by Chiang \textit{et al.}\cite{Chiang.2020}.
Due to coarse geometric features compared to the wavelength, acoustic metagrating are scalable to low-frequency ultrasound range, as demonstrated by Chiang \textit{et al.} \cite{Chiang.2021} with a \unit[40]{kHz} design.%

Microacoustic metagratings consist of micro-fabricated structures, the natural consequence of this microscale being the significant influence of thermoviscous effects. 
The viscosity begins to strongly contribute when the Stokes's boundary layer thickness $\delta_{\mathrm{S}} = 2 \pi \sqrt{2 \mu / (\omega \rho_0)}$ being related to the well known viscous boundary layer $\delta_{\mathrm{v}} = \sqrt{\mu/(\omega \rho_0)}$ as $\delta_{\mathrm{S}} = 2 \pi \sqrt{2} \delta_{\mathrm{v}}$ approaches the geometrical dimensions of the channels \cite{Ward.2015, Cotterill.2018}.
In metagratings, the dimensions of the meta-atoms are close to the wavelength $\lambda = c_0/f$, which, however, decreases faster with increasing frequency than the thickness of the boundary layers.
Therefore we introduce a normalized Stoke's boundary layer thickness
\begin{equation}\label{eq:betas}
    \beta_{\mathrm{S}} = \frac{\delta_{\mathrm{S}}}{\lambda} = \frac{ \sqrt{f}}{c_0 }\sqrt{\frac{4 \pi \mu}{\rho_0}} 
\end{equation}
that relates the boundary layer thickness to the wavelength and then indirectly to the metagrating dimensions.
Accordingly, the thermal boundary layer
$\delta_{\mathrm{t}} = \sqrt{k / (2 \pi f \rho_0 C_p ) }$ being linked to $\delta_{\mathrm{v}}$ by the Prandtl number as $\delta_{\mathrm{t}} = \delta_{\mathrm{v}} / \sqrt{\mathrm{Pr}}$ \cite{Ward.2015, Cotterill.2018}
can be normalized as
\begin{equation}\label{eq:betat}
    \beta_{\mathrm{t}} = \frac{\sqrt{f}}{c_0}\sqrt{\frac{k}{2 \pi \rho_0 C_p}}.
\end{equation}
Fig.~\ref{fig:meta_basic}b illustrates the normalized Stoke's boundary layers $\beta_{\mathrm{S}}$ over the frequency, where it can be observed that a change of two order of magnitude in frequency (scaling from audio range to ultra-high frequency ultrasound) leads to an order of magnitude change in boundary layer thickness.

To demonstrate the current state of the art, Fig.~\ref{fig:meta_basic}b plots the gradient metasurfaces and metagratings reported in the literature, showing the relation of the thinnest channel dimension $l_{\mathrm{min}}$ to $\delta_{\mathrm{S}}$.
We note that previously reported metamaterial structures from the literature in Fig.~\ref{fig:meta_basic}b are limited by \unit[40]{kHz}.%
In the case of gradient metasurfaces (Fig.~\ref{fig:meta_basic}b, blue markers), the ratio $\delta_{\mathrm{S}} / l_{\mathrm{min}}$ ranges from \unit[6]{\%} to \unit[25]{\%}, where thermoviscous effects can be expected.
If however, these gradient metasurface designs would be scaled to ultra-high frequency ultrasound (shifting in parallel to the $\beta_{\mathrm{S}}$ curve), in most cases the Stoke's boundary thickness would exceed the channel width (Fig.~\ref{fig:meta_basic}b, black dotted line) significantly changing the designed phase and leading to high absorption.
Metagratings offer a huge potential to reduce losses by using a channel width close to the wavelength (Fig.~\ref{fig:meta_basic}b, red markers).
Nevertheless, scaling to ultra-high frequency ultrasound is still a challenge, since the Stoke's boundary thickness is close to \unit[10]{\%} of the channel width, where thermoviscous effects need to be taken into account in the design process.

Here we report for the first time on the largerly unexplored concept of microacoustic metagrating in airborne ultra-high frequency ultrasound, as a route to use acoustic metamaterials for flat acoustics, ultrasound imaging, and acoustic spectroscopy with high spatial and temporal resolution, where it is necessary to operate at ultra-high frequency ultrasound.
The target frequency in this work is set to \unit[2]{MHz}, which is far above the frequency range previously discussed in literature (see Fig.~\ref{fig:meta_basic}b).
We choose $\lambda$ and $d$ in a range in which only three propagating diffraction orders ($-1$, $0$, and $+1$) are supported, for both refracted and reflected waves.
With a view to demonstration of a diffraction angle that allows a sufficiently short focal length for flat acoustic lenses, for example, the target angle is set to $\theta_{-1} = \unit[-35.0]{^\circ}$, while the wave is incident at $\unit[0]{^\circ}$.
The resulting lattice constant of the metagrating $d = \unit[299]{\mu m}$ is given by the Bragg's condition, such that the $-1$st diffraction order is aligned with the desired angle.
As a result, there are six different directions in which the wave can be redirected by the metagrating, three transmitted ($T_{-1}$, $T_{0}$, and $T_{+1}$) and three reflected ($R_{-1}$, $R_{0}$, and $R_{+1}$) diffraction orders.
Because of $ \beta_{\mathrm{S}} \approx 10^{-1} $, we expect considerable viscous losses that open an additional (seventh) energy channel, namely absorption $ \alpha $.
In order to design meta-atoms for microacoustic metagratings, we apply the linearized Navier-Stokes equations implemented within the finite element method (FEM) framework. 
We use different design strategies, all of which are finalized through FEM-based shape optimization including thermoviscous effects, maximizing the energy transmitted into the $ -1 $st diffraction order.
Finally, we design and manufacture three different microacoustic metagratings using two-photon polymerization.
This manufacturing process has not yet been explored for acoustic metamaterials and we demonstrate for the first time its application to acoustic metagratings.
Unidirectional acoustic beams achieved by our advanced designs are attributed to the bianisotropic properties of the optimized metaatoms and to the improvement of the models compared to the commonly used lossless approaches.
Although part of the acoustic energy is dissipated due to the unavoidable effect of viscous losses in the narrow slits, our numerical and experimental results demonstrate that highly directional refraction can be achieved.
The metagrating structures presented can be used for most ultrasound applications, for example flat acoustic lenses for ultrasound imaging with a high temporal and spatial resolution or the directional shaping of high-frequency ultrasonic transducers for gesture recognition.

\section{Results}

\subsection{Can thermoviscous effects be neglected?}

\begin{figure}
    \centering
    \textbf{a}\includegraphics[height = 4.0cm]{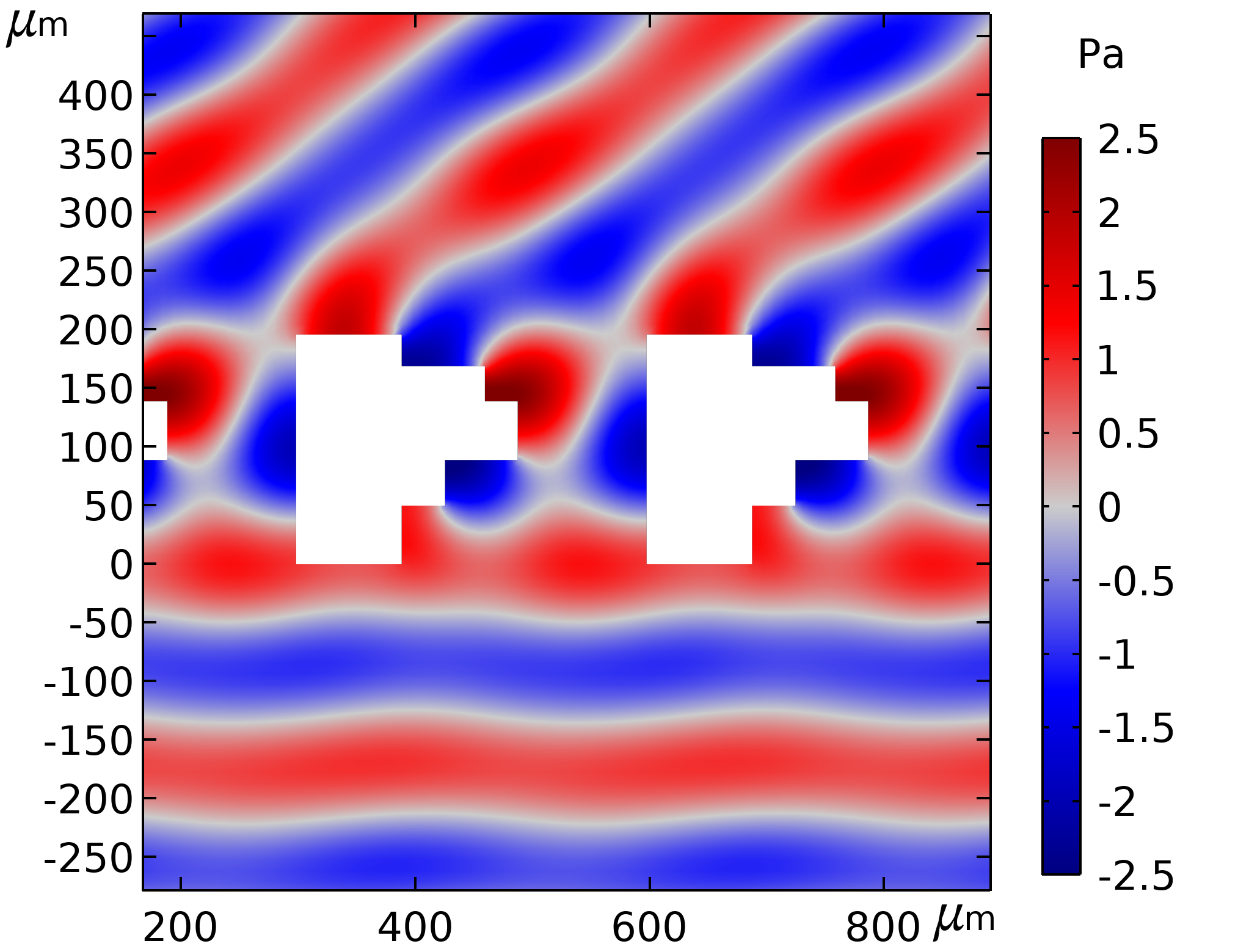}
    \textbf{b}\includegraphics[height = 4.0cm]{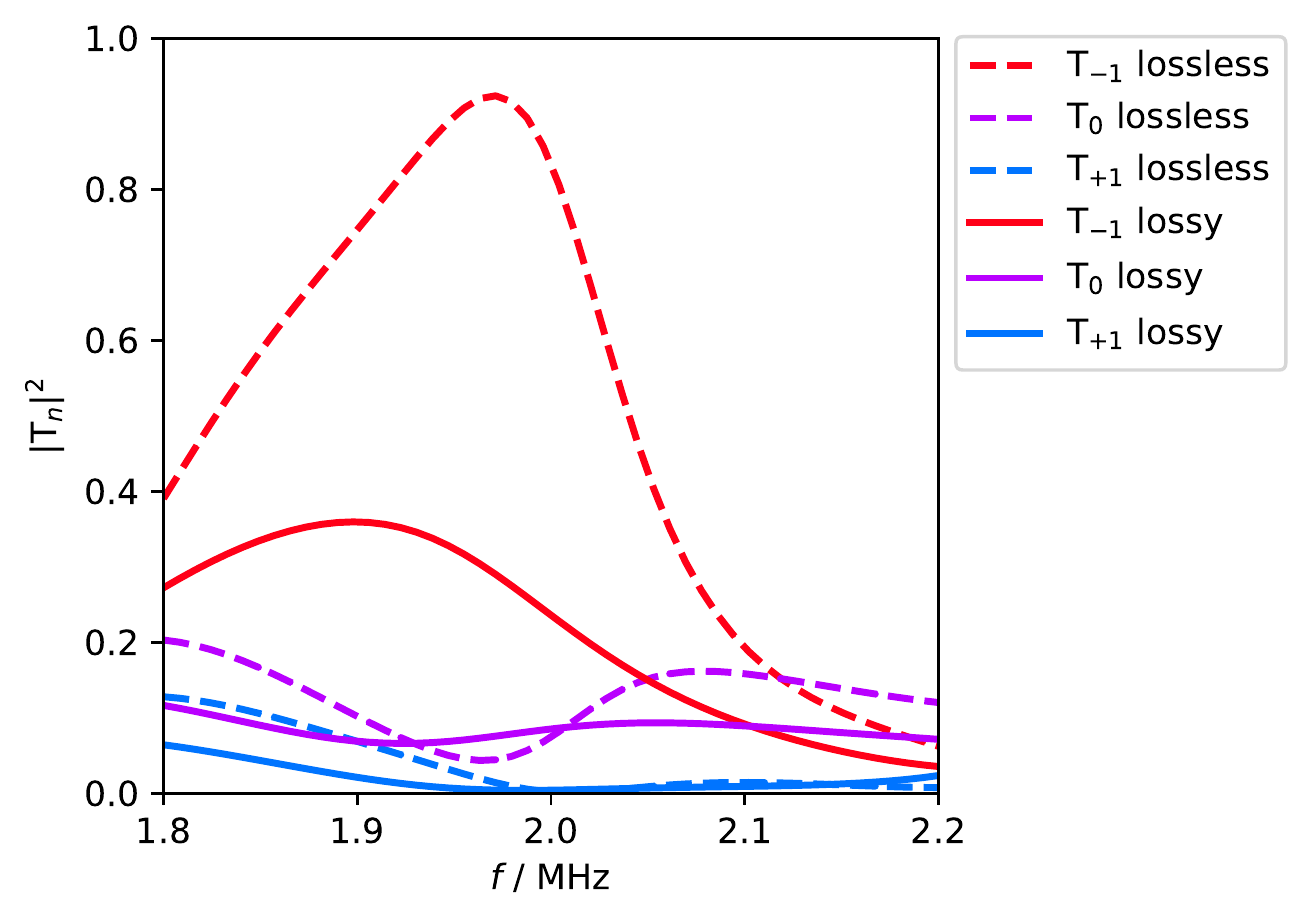} \\
	\textbf{c}\includegraphics[height = 4.0cm]{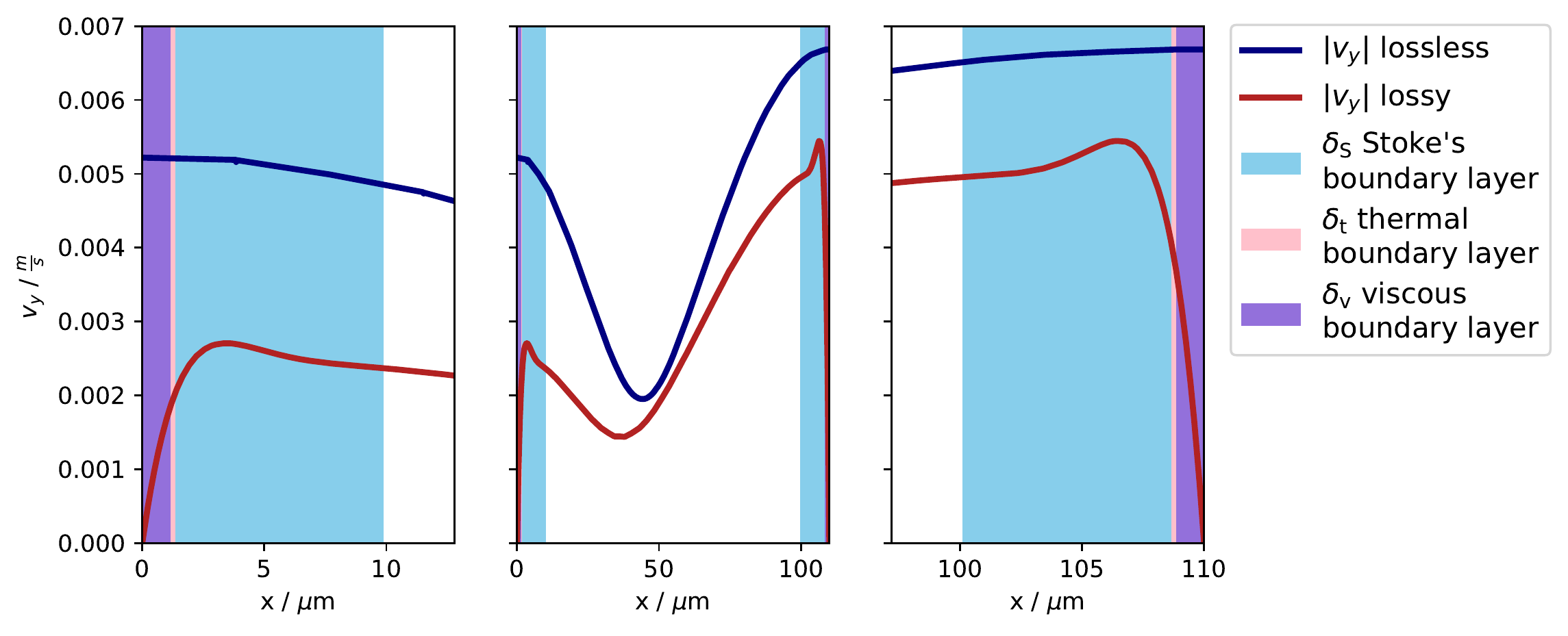}\\
    \textbf{d}\includegraphics[height = 4.5cm]{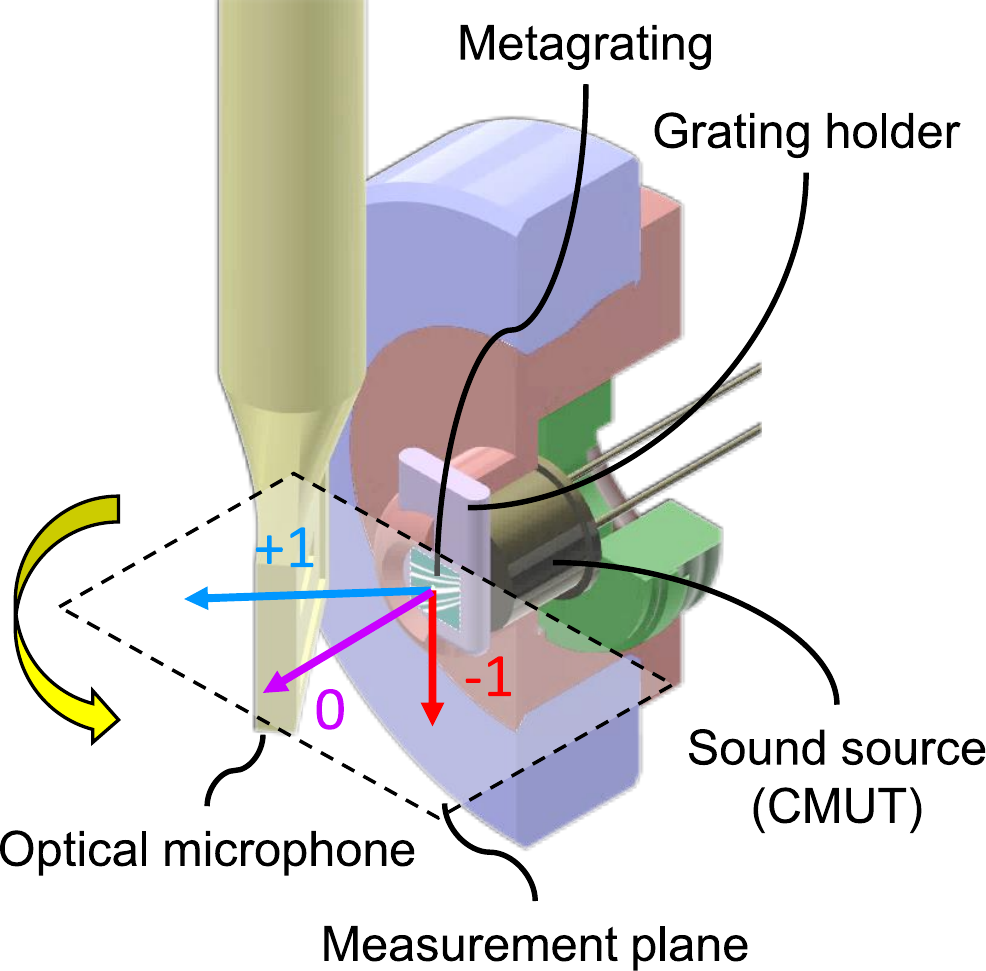}
    \textbf{e}\includegraphics[height = 4.5cm]{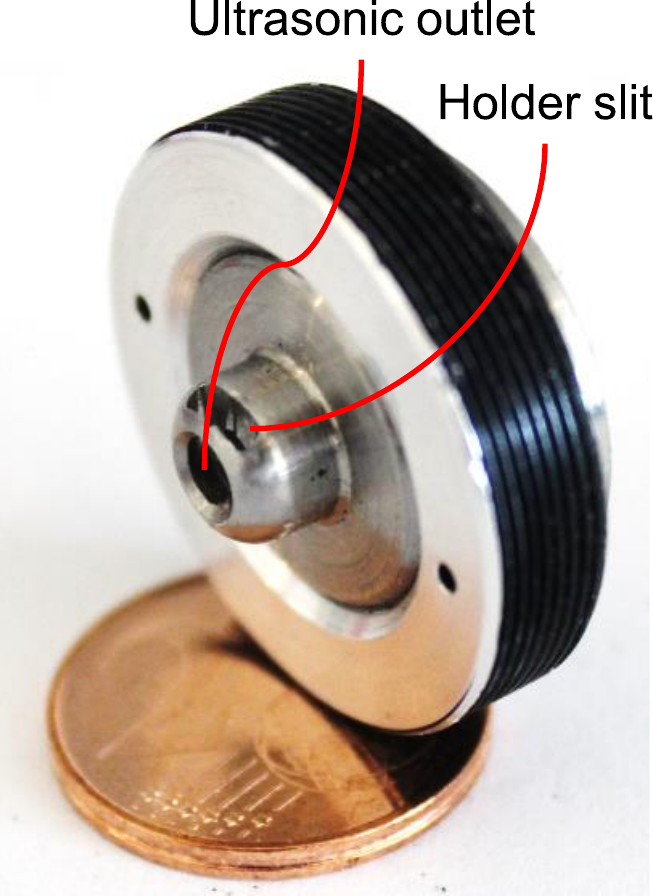}\\
    \textbf{f}\includegraphics[height = 4.5cm]{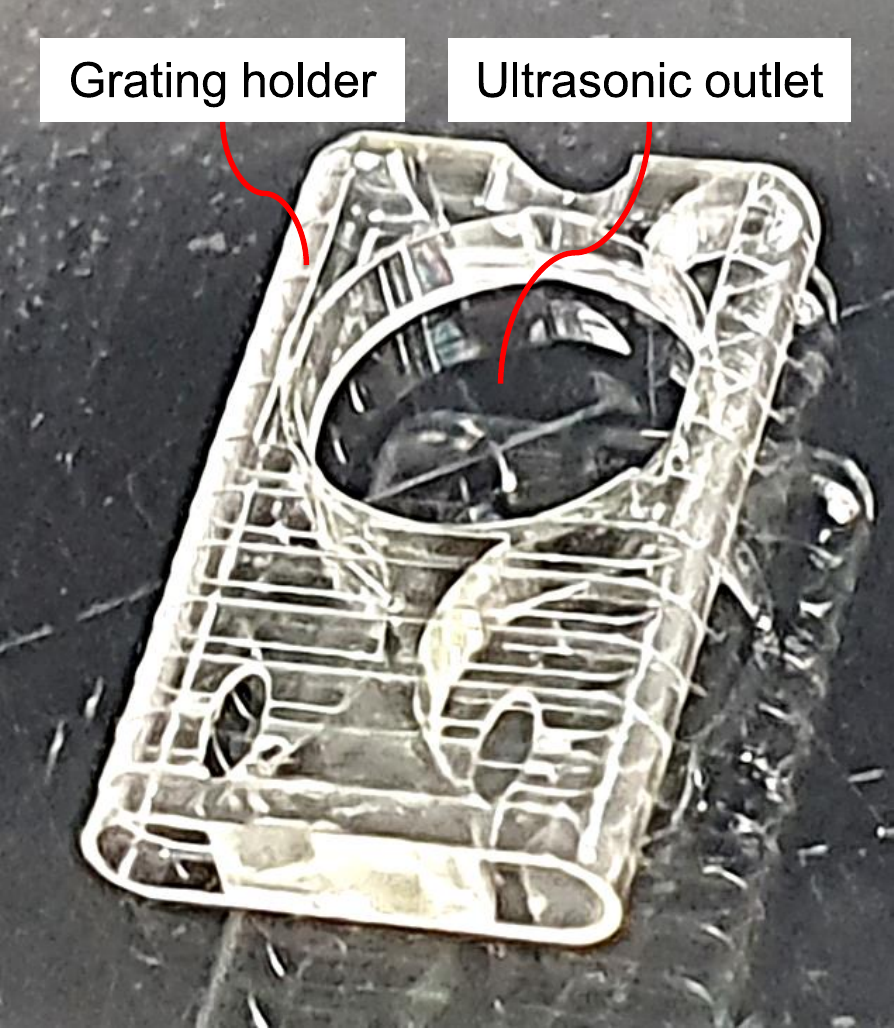}
    \textbf{g}\includegraphics[height = 4.5cm]{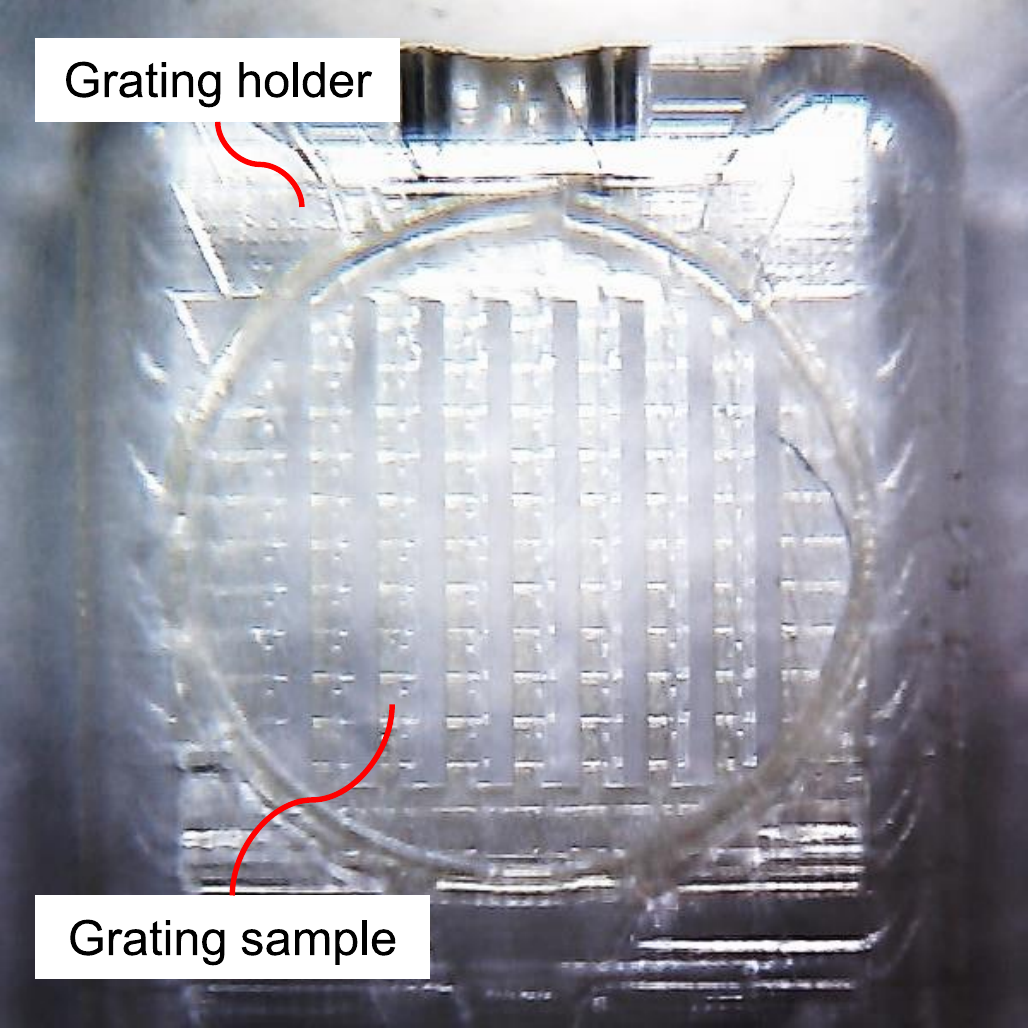}
    \caption{
        \textbf{Lossless metagrating design and experimental setup}
        \textbf{a} Real part of the pressure at the peak frequency in lossless case.
        \textbf{b} Transmission coefficients for the lossless model and the model with thermoviscous effects.
        \textbf{c} Comparison of the $y$ velocity at \unit[2]{MHz} in the narrowest slit between two metaatoms of design A. The right and left panels show a detailed view of the curves close to the boundary.
        \textbf{d} The rotatable holder including transducer and the moving laser microphone.
        \textbf{e} Photograph of the rotatable holder.
        \textbf{f} Micrograph of the grating holder created by two-photon lithography.
        \textbf{g} Micrograph of the microacoustic metagrating sample inside the grating holder.
    }
    \label{fig:D0_and_exp_setup}
\end{figure}

The first metagrating (design A) is designed based on the Helmholtz equation, which does not consider the viscosity of air and is evolved from a two-dimensional model assuming an infinite number of meta-atoms.
For the optimization of lossless structures, simple geometrical shapes are used in combination with a semianalytical model based on the superposition of waveguide modes similar to Ref.~\citenum{Chiang.2021}.
The final shape of the metaatom consists of five rectangles, which add together to a thickness $t = \unit[196]{\mu m}$, see Fig.~\ref{fig:D0_and_exp_setup}a.
This design promises $|T|^2 \approx 0.95$ around \unit[2]{MHz} when no losses are considered, where the numerical results are shown in Fig.~\ref{fig:D0_and_exp_setup}b as dashed line with $|T|^2$ in red.
However, when the more accurate linearized Navier-Stokes equations are used, transmission towards $-1$st diffraction order is strongly reduced, while the peak shifts to \unit[1.90]{MHz} with only $|T_{-1}|^2 = 0.36$.
Furthermore, we observe a redistribution of the transmitted energy to the other diffraction orders and not just a reduction of all magnitudes due to losses as might be expected.
The transmission towards 0th order is even bigger as in the lossless case between \unit[1.94]{MHz} and \unit[2.00]{MHz}, which means that the design requires further optimization.

In order to review the differences between both modelling strategies, the harmonic fluid velocity in y-direction at the narrowest location between two meta-atoms is shown in Fig.~\ref{fig:D0_and_exp_setup}c.
In the lossy case (red curves) the velocity is reduced compared to the lossless case (blue curves), while it is zero at the boundary. Furthermore the entire profile deviates from the lossless case, especially within the Stoke's boundary layer (light blue region).
It should be noted, that the assumption of the zero velocity at the boundary is only valid for Knudsen number $\mathrm{Kn} < 0.01$, which means a channel width $< \unit[7]{\mu m}$.
In contrast to the viscous effects, the observed shift of the $|T_{-1}|^2$ peak towards lower frequencies could be linked to the thickness of the thermal boundary layer $\delta_{\mathrm{t}}$, despite being only around \unit[1]{\%} of the channel width.
Similar effects are known from narrow slits \cite{Ward.2015} and from resonance based bianisotropic acoustic meta-atoms \cite{Quan.2018, Melnikov.2019}.
However, it is still unclear whether the reduction in transmission and the shift in the peak frequency can be resolved experimentally.

To characterize the designed and fabricated microacoustic metagratings we use the experimental setup shown in Fig.~\ref{fig:D0_and_exp_setup}d combining a grating mounting with an ultra-high frequency capacitive micromachined ultrasonic transducer with a center frequency of \unit[2]{MHz} (see supplementary materials), a laser microphone and precise motorized stages \cite{Koble.2021}.
For experimental realization, a finite number of 9 meta-atoms is extruded into the third dimension $z$ and placed at the end of the ultrasonic aperture with a diameter of \unit[3]{mm}.
While the transducer emits a plane wave towards the metagrating through the ultrasound aperture, the microphone detects the transmitted waves as pressure oscillations at different angles and radii in the $xy$-plane located in the middle of the metagrating. The mounting includes a rotatable holder (Fig.~\ref{fig:D0_and_exp_setup}e), the grating holder (Fig.~\ref{fig:D0_and_exp_setup}f), and the metagrating sample (Fig.~\ref{fig:D0_and_exp_setup}g).
The metagrating samples are manufactured by two-photon polymerization, while the mechanical properties of the resulting polymer IP-S are sufficient to create an impedance contrast close to a hard boundary for airborne sound.

Figure~\ref{fig:all_results}a shows a scanning electron microscope (SEM) image of the cross section of the microscopic metagrating manufactured by two-photon lithography. 
We define the normalized transmission
\begin{equation}\label{eq:tau}
    \tau_n = \frac{|T_{n}|^2}{\sum_{m} |T_{m}|^2}, \, m \in \left\{0, \pm 1 \right\}
\end{equation}
to characterization the performance of the metagrating.
The measured and numerically calculated normalized transmission is shown in Fig.~\ref{fig:all_results}b, while the simulations of the unnormalized transmission are included in the supplementary material.
Here we observe how accurately the experimental results match the simulation with thermoviscous effects.
The numerically determined maximum $\tau_{-1}$ peak is located at \unit[1.93]{MHz} with $\tau_{-1} = 0.82$, while in the experiment we observe the peak at \unit[1.92]{MHz} with $\tau_{-1} = 0.82 \pm 0.02$.
In addition, the numerically predicted crossing point of $\tau_{-1}$ and $\tau_0$ around \unit[2.10]{MHz} is also present in the experiment.
From the simulation, we observe that half of the energy is absorbed in the metagrating around the target frequency (see Fig.~\ref{fig:all_results}c, black dashed line), while the maximum absorption is located at \unit[1.98]{MHz} with $\alpha = 0.58$.
The measured directivity profile at the target frequency \unit[2]{MHz} is shown in Fig.~\ref{fig:all_results}c, where we see strong refraction towards the intended angle $\theta_{-1} = \unit[-35]{^{\circ}}$.
One option to increase the performance is to use the peak frequency at \unit[1.92]{MHz}, which is \unit[4]{\%} below the target frequency, but this would change the diffraction angle to $\theta_{-1} \approx \unit[-37]{^{\circ}}$.
However, this is an unacceptable compromise for flat acoustic lenses consisting of several different metagrating areas, for example, since it would lead to defocusing.
Therefore, we conclude that thermoviscous effects must not be neglected during the design of microacoustic metagratings.

\subsection{Reoptimization including thermoviscous effects (design B)}

To overcome the drop in efficiency and frequency downshift of the peak, the design is shape reoptimized using FEM including thermoviscous losses based on linearized Navier-Stokes equations.
During the optimization the free shape boundary method \cite{comsol} is used, while the shape deformation was limited to $\lambda/4$ from initial boundary location.
The lossless design is used as the initial geometry, while the metagrating thickness $t$ is fixed.
The optimization target is set to maximize the transmission toward $-1$st diffraction order and the maximum value achieved is $|T_{-1}|^2 = 0.45$.
The final geometry of design B manufactured using two-photon polymerization and is shown in Fig.~\ref{fig:all_results}d.
It can be seen that the optimal structure resembles the initial design A and most of the geometric features are maintained.

The comparison between numerically and experimentally determined normalized transmission is shown in Fig.~\ref{fig:all_results}e. 
The experimental result matches the numerics with high accuracy considering the peak and the crossings.
From the numerics we observe the peak at \unit[2.02]{MHz} with $\tau_{-1} = 0.87$.
This is confirmed by the experimental result with $\tau_{-1} = 0.83 \pm 0.03$ at \unit[2.02]{MHz}.
In addition, in FEM, as well as in the experiment $\tau_{-1}$ and $\tau_0$ come very close to each other around \unit[1.80]{MHz} and \unit[2.20]{MHz}.
Compared to initial design A, the model suggests that the absorption coefficient $\alpha$ is reduced in the investigated frequency range, while it is particularly observable around \unit[2]{MHz}, where the absorption coefficient $\alpha$ drops from 0.57 to 0.44.
At the target frequency there is a small performance drop (see Fig.~\ref{fig:all_results}e), which cannot be explained by the numerical model, but makes operation at the target frequency slightly less attractive as at the neighboring frequency steps.
Subsequently, viewing the measured directivity pattern at the target frequency would give a result similar as for design A.
Therefore, Fig.~\ref{fig:all_results}f shows the directivity at \unit[2.02]{MHz} (\unit[1]{\%} above the target, $\theta_{-1} \approx \unit[-34.6]{^{\circ}}$) with $\tau_{-1}=0.83$ being an improvement compared to design A with $\tau_{-1}=0.82$ at \unit[4]{\%} lower frequency.%

\subsection{Broadband and subwavelength microacoustic metagrating (design C)}

To tackle the demand for broadband and subwavelength metagratings, we introduce a new design C, which is the result of a combination of topology optimization with shape optimization.
Furthermore, thinner metagratings in general have shorter channel length, which can help to reduce unwanted absorption.
In the first step a new geometry is created by topology optimization aiming for maximum $T_{-1}$ without the consideration of thermoviscous losses.
The optimization domain is a wall of infinite length and a thickness $t = \lambda / 4$, which subsequently is the thickness limit in that step.
In the second step, the newly created geometry is taken as the initial geometry for a shape optimization under consideration of thermoviscous losses.
Again the free shape boundary method \cite{comsol} with deformation limitation of $\lambda/4$ is used.
The final geometry promises $|T_{-1}|^2 = 0.41$ at the target frequency with a thickness of only $t=\unit[50]{\mu m} \approx 0.29 \lambda$.
It should be noted that the resulting meta-atom consists of two separate bodies, which is surprising since it was not specified.
To the best of our knowledge, such a configuration is new and unexplored in the context of acoustic metagratings.

\begin{figure}
    \begin{tabular}{cccc}
	\textbf{a} \includegraphics[height = 4.0cm]{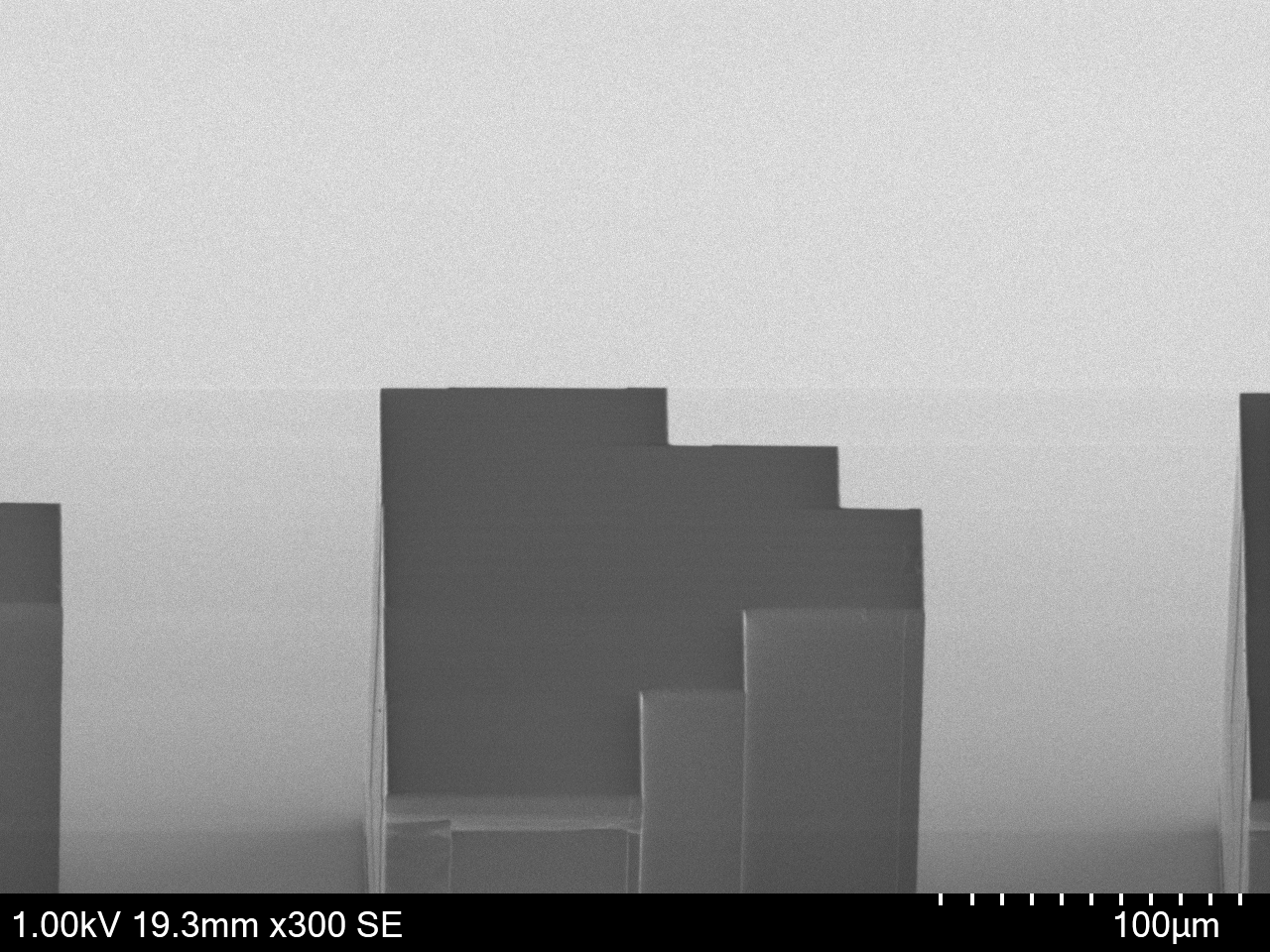} &
	\textbf{b} \includegraphics[height = 4.0cm]{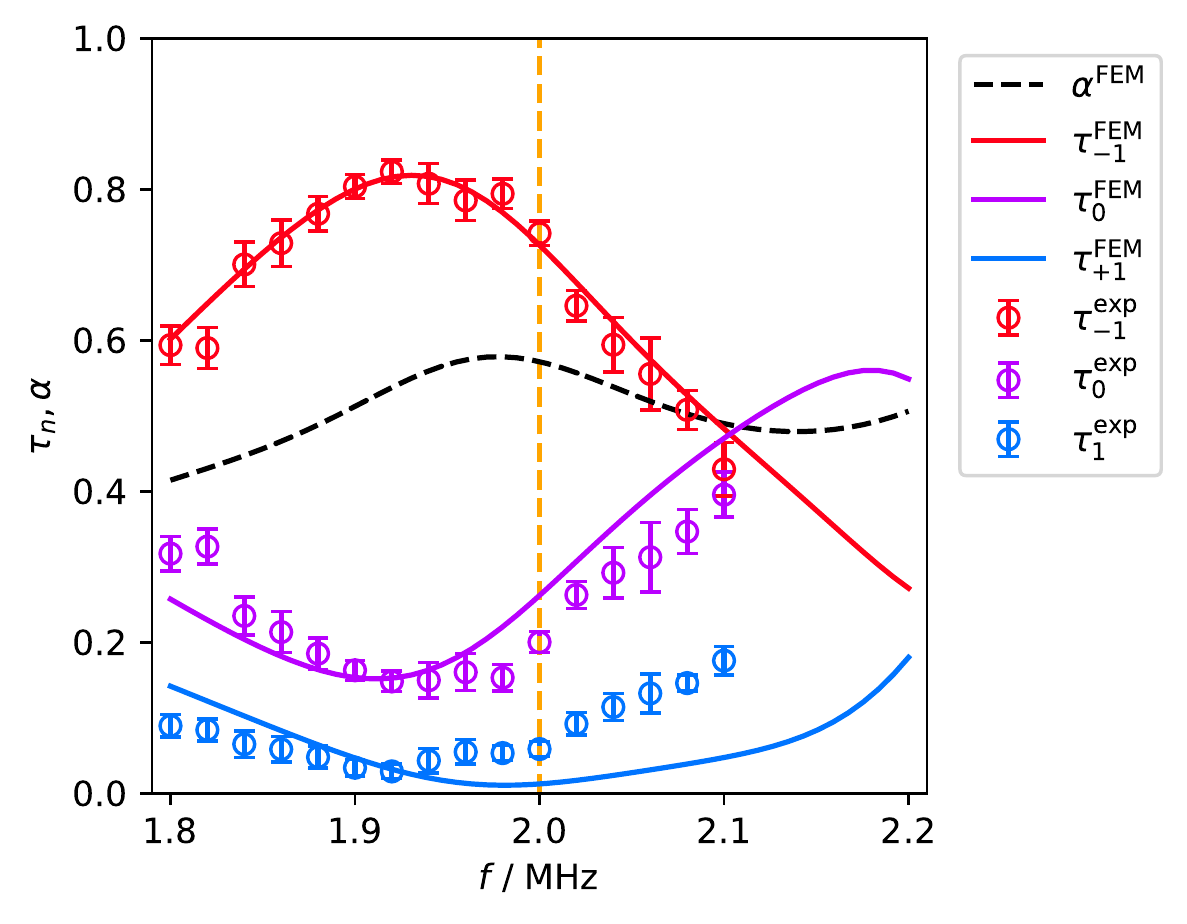}
	\textbf{c} \includegraphics[height = 4.0cm]{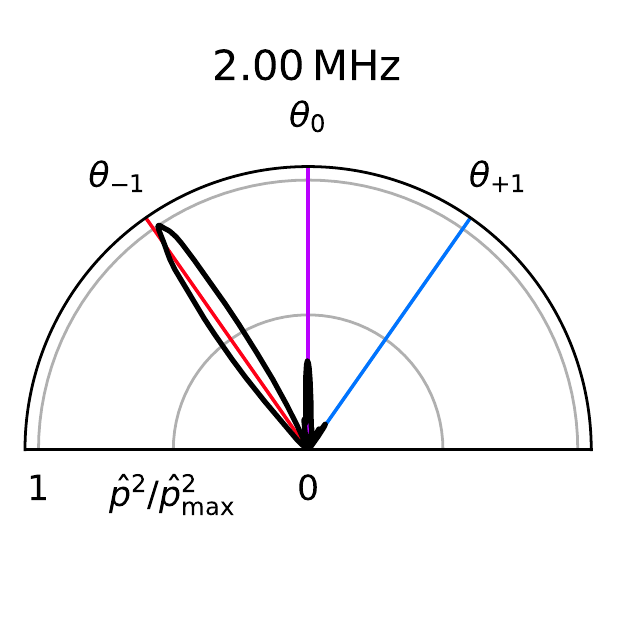}\\
	\textbf{d} \includegraphics[height = 4.0cm]{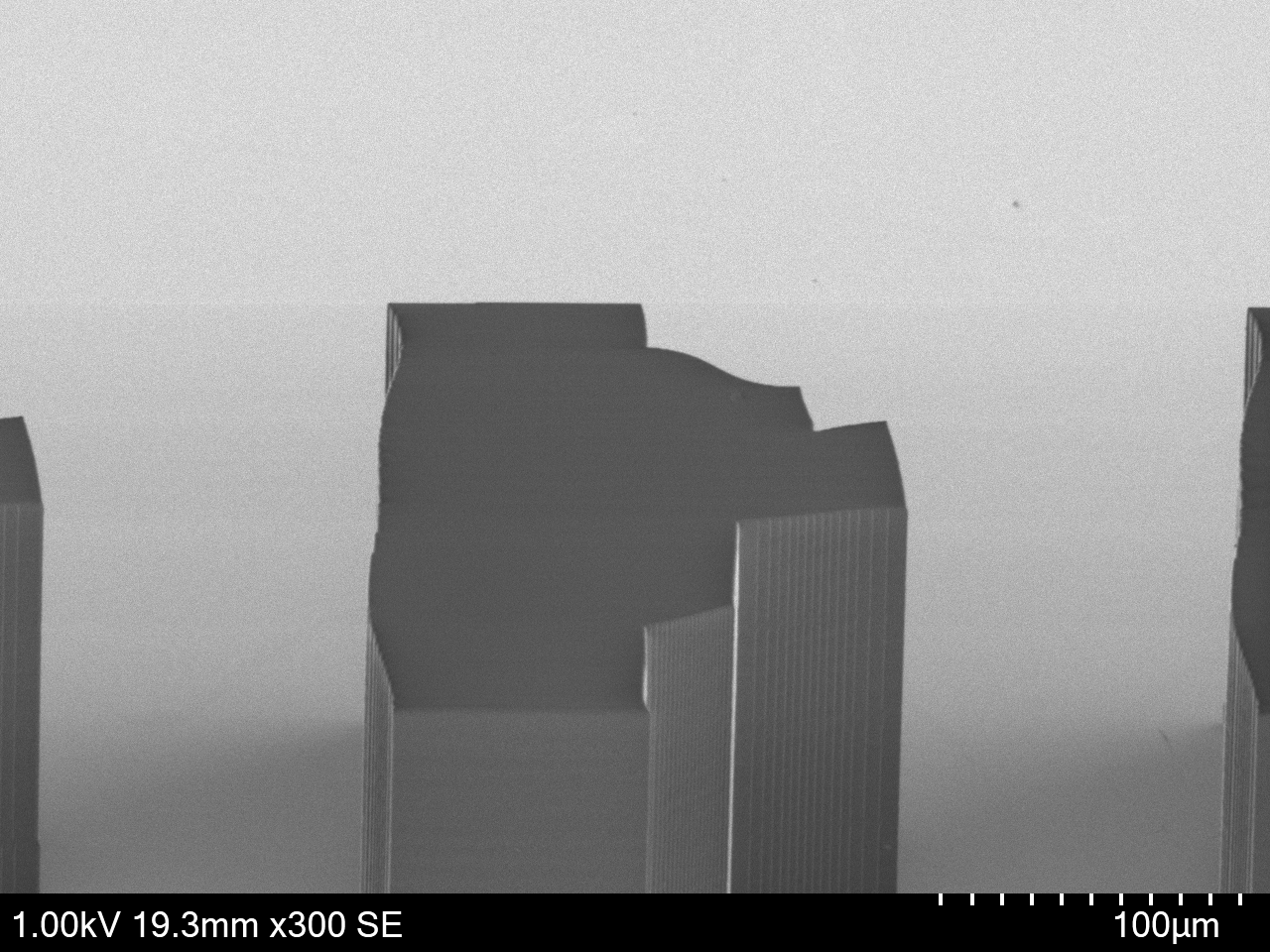} &
	\textbf{e} \includegraphics[height = 4.0cm]{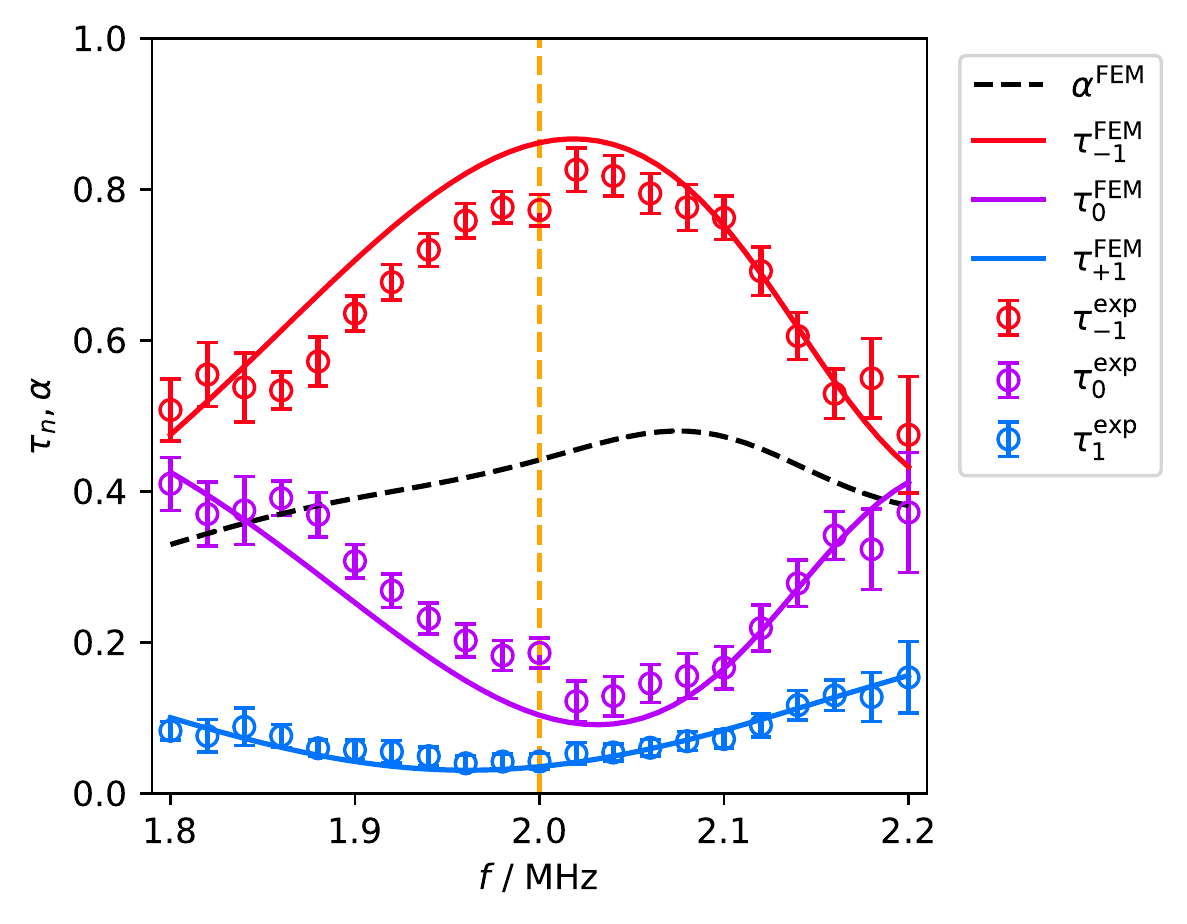}
	\textbf{f} \includegraphics[height = 4.0cm]{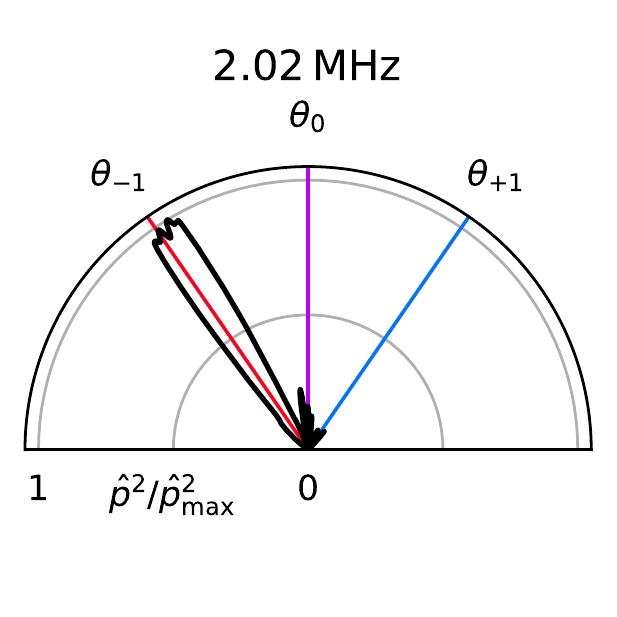}\\
	\textbf{g} \includegraphics[height = 4.0cm]{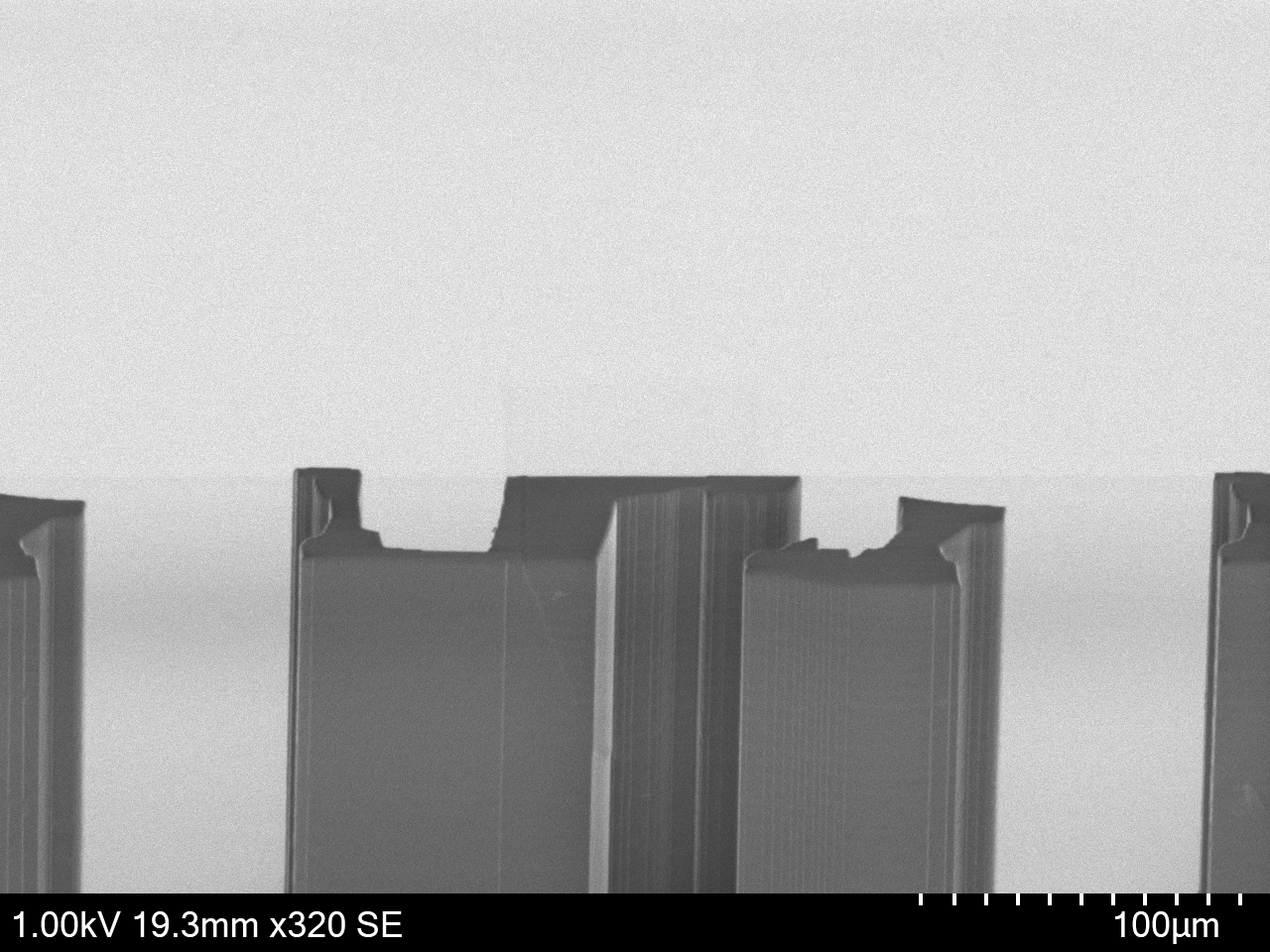} &
	\textbf{h} \includegraphics[height = 4.0cm]{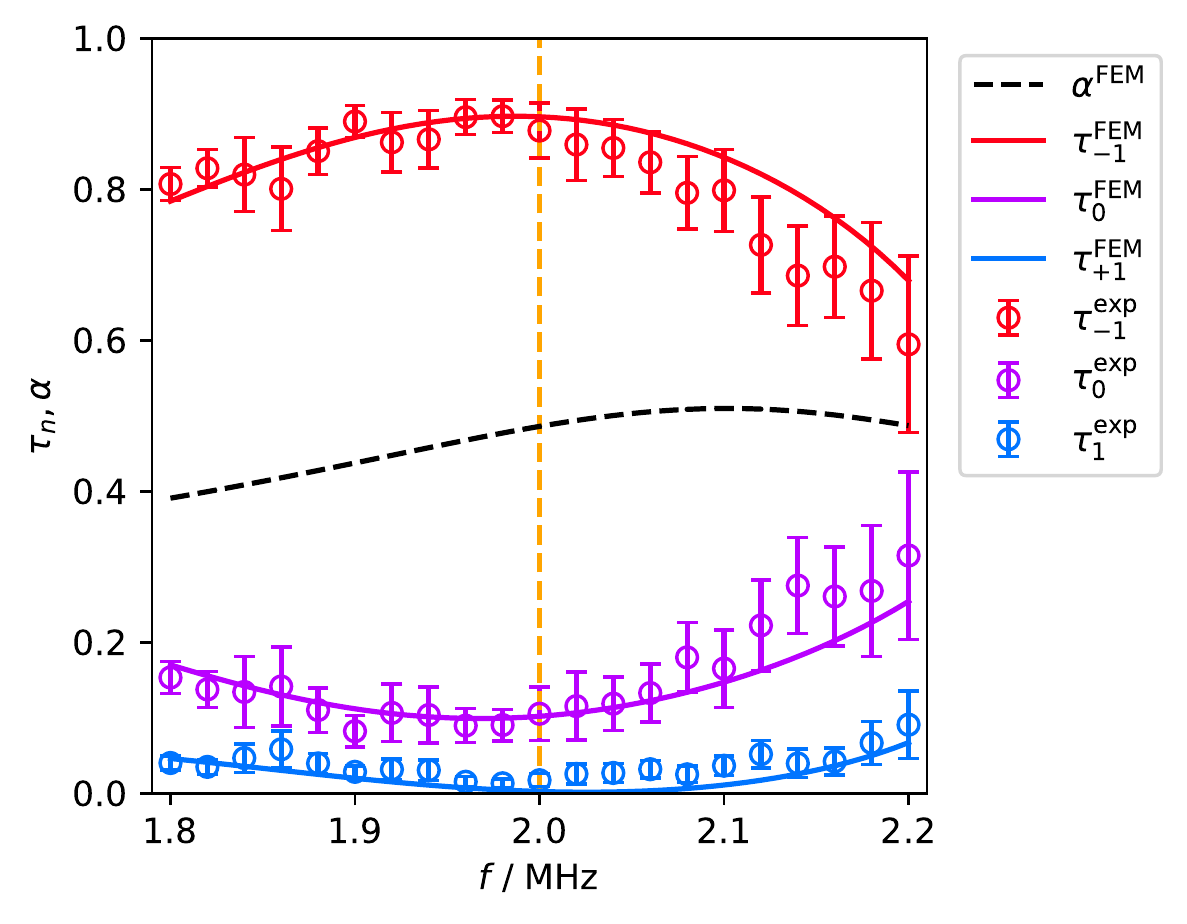}
	\textbf{i} \includegraphics[height = 4.0cm]{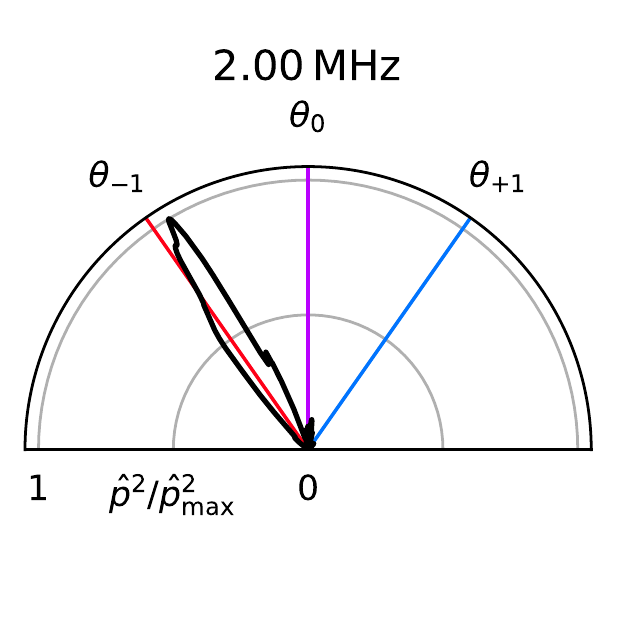}\\
    \end{tabular}
	\caption{
	    \textbf{Designed and manufactured microacoutic metagratings and experimental results.}
	    \textbf{a-c} Design A, semi-analytically optimized lossless metagrating.
	    \textbf{d-f} Design B, reoptimized metagrating including thermoviscous effects.
	    \textbf{g-i} Design C, broadband and subwavelength microacoustic metagrating.
	    \textbf{adg} SEM micrograph of the meta-atom manufactured by two-photon lithography.
	    \textbf{beh} Experimentally and numerically determined normalized transmission $\tau_n$ and numerically determined absorption $\alpha$.
	    \textbf{cfi} Directivity pattern around the target frequency.
	}
    \label{fig:all_results}
\end{figure}

Figure~\ref{fig:all_results}g shows an SEM image of the manufactured metagrating. The shape includes thin walls and jagged edges, which are accurately recreated by two-photon lithography.
The experimentally and numerically determined transmission are shown in Fig.~\ref{fig:all_results}h, where we observe quite high amount of energy in the $-1$ diffraction order.
From the numerics the $\tau_{-1}$ peak is expected at \unit[1.98]{MHz} with $\tau_{-1} = 0.90$, while in the experiment the peak is located at \unit[1.96]{MHz} with $\tau_{-1} = 0.90 \pm 0.02$. 
In addition, $\tau_{-1} = 0.88 \pm 0.04$ at the target frequency, being better than the peak values of previous designs A and B.
Furthermore, we note that in the frequency range between \unit[1.80]{MHz} and \unit[2.08]{MHz} $\tau_{-1} > 0.80$, which is broadband compared to previous results.
The absorption (see Fig.~\ref{fig:all_results}h, black dashed line) with $\alpha = 0.49$ at the target frequency is slightly higher than in the design B (see Fig.~\ref{fig:all_results}e), but still much lower than in the design A.
Taking a look at the directivity pattern in Fig.~\ref{fig:all_results}i reveals strong diffraction towards $-1$st order with other orders being negligible.
Although this new design has thinner channels with dimensions closer to the Stoke's boundary layer thickness $\delta_{\mathrm{S}}$ (see Fig.~\ref{fig:meta_basic}b, green cross), it demonstrates the best experimental performance from the investigated geometries.

\section{Discussion}

How could the unexpected better performance of the subwavelegth design C be explained, when the range of the possible shapes is significantly limited due to the bounded thickness?
This is due to a different topology of the meta-atom consisting of two separate bodies, which was found through the naivety of the topology optimization.
We can gain further insight by looking at this from the perspective of the fluidic channels, where, contrary to all previously reported acoustic metagrating designs, here we have two channels instead of a single one within a period.
The dimensions of these channels are different, resulting in a transmission phase shift between the channels that is comparable to gradient metasurfaces.
We note that for a gradient metasurface, two channels is the absolute minimum to break symmetry, however typically more are required to match the required phase distribution.
Bianisotropic metaatoms consisting of multiple separated bodies have been already discussed in the optics \cite{Rahmanzadeh.2020} and in the acoustics, although limited to multiple solid cylinders \cite{Fan.2020, Fan.2021} or to a cylindrical shape with a resonant cavity \cite{Quan.2018, Melnikov.2019}.

In the current work we have demonstrated the microacoustic metagratings in air, but the aforementioned applications are at least as important for waterborne ultrasound.
Using the normalized expressions for boundary layer thicknesses in Eqs.~\eqref{eq:betas} and \eqref{eq:betat} our results can be transferred to other media, particularly water.
Considering water with
dynamic viscosity $\mu^{\mathrm{water}} = \unit[1.0093 \times 10^{-3}]{Pa \cdot s}$,
speed of sound $c^{\mathrm{water}} = \unit[1418.1]{m \cdot s^{-1}}$,
and equilibrium density $\rho_0^{\mathrm{water}} = \unit[998.2]{kg \cdot m^{-3}}$
the current airborne study at \unit[2]{MHz} is similar to a waterborne study at $\unit[559]{MHz}$ with a grating constant of $d = \unit[4.63]{\mu m}$ from the view of viscosity.
However, the normalized thermal boundary layer of thickness $\beta_{\mathrm{t}}$ has also non-negligible effects, leading to a second similarity parameter $\beta_{\mathrm{t}}$ linked by the Prandtl number Pr to $\beta_{\mathrm{S}}$.
Since the Prandtl number of water $\mathrm{Pr}^{\mathrm{water}} = 7.1$ (thermal conductivity $k^{\mathrm{water}} = \unit[0.59423]{W \cdot m^{-1} \cdot K^{-1}}$ and heat capacity at constant pressure $C_p^{\mathrm{water}} = \unit[4186.9]{J \cdot kg^{-1} \cdot K^{-1}}$) is one order of magnitude larger than that of air $\mathrm{Pr}^{\mathrm{air}} = 0.71$, the normalized thermal boundary layer $\beta_{\mathrm{t}}$ is smaller and, from the point of view of $\beta_{\mathrm{t}}$, the present study is comparable to \unit[5.6]{GHz} with $d = \unit[461]{nm}$ in water.
Even if thermal and viscous effects in our numerical model cannot be separated, the viscosity is the dominant source of the losses and $\beta_{\mathrm{S}}$ is the more relevant similarity parameter.
From this we conclude that our results can be interpreted to expect a feasibility of microacoustic metagratings in water above $\unit[500]{MHz}$, which are considered to be deep ultra-high frequency for waterborne ultrasound.
This holds only, when the fabricated metagrating material reveals an impedance contrast close enough to sound-hard boundary, which is not the case when the same polymers as in the current work are used.
This only applies if the metagrating material produced has an impedance contrast comparable to the sound-hard boundary, which is not the case when using the same polymers as in the current work.
This can be solved by manufacturing glasses with two-photon lithography \cite{Kotz.2021}, otherwise the structural dynamics must be included in the design process via the fluid-structure interface.

\section{Conclusions}

We have demonstrated the feasibility of metagratings fabricated by two-photon polymerisation for control of ultra-high frequency ultrasound.%
We proved experimentally that anomalous refraction at ultra-high frequencies in air ($\geq \unit[2]{MHz}$) is possible, which is of particular interest for flat acoustics and ultrasonic imaging  with  high  spatial and temporal  resolution.
For design purposes it is essential to include thermoviscous effects, since the boundary layer thickness becomes close to the geometric dimensions at such high frequencies.
These results push forward research and innovation in the field of ultra-high ultrasound waves in air and other viscous fluids.

\section{Methods}

\subsection{Numerical model}
For numerical solution the linearized Navier-Stokes model was solved using the COMSOL Multiphysics software. An infinite metagrating in two-dimensional space was assumed and therefore only a single strip of the length $d$ was modelled. To ensure the periodicity of the solution a Bloch-Floquet boundary condition was applied. The infinite domain extension was modelled by using perfectly matched layers (PML) above and below the metagrating.

In order to obtain plane wave expansion coefficients a surface $A$ of the size $d \times d$ was defined before (reflection, $R$) and after (transmission, $T$) the metagrating leading to
\begin{equation}
\begin{aligned}
    a_n^T &= \frac{1}{A_T} \int_{A_T} p(\mathbf{r}, f) e^{i \mathbf{k}_n^T \mathbf{r}} \textrm{d}A_T \\
    a_n^R &= \frac{1}{A_R} \int_{A_R} p(\mathbf{r}, f) e^{i \mathbf{k}_n^R \mathbf{r}} \textrm{d}A_R
    \\
    a_{\mathrm{inc}} &= \frac{1}{A_R} \int_{A_R} p(\mathbf{r}, f) e^{i \mathbf{k}_{\mathrm{inc}} \mathbf{r}} \textrm{d}A_R
\end{aligned}
\end{equation}
with $n \in \left\{0, \pm 1 \right\}$ being the diffraction order, $p(\textbf{r}, f)$ being the complex valued pressure, and $\textbf{r}$ being the location vector.
The wave vector $\textbf{k}$ is defined as
\begin{equation}
\begin{aligned}
    \textbf{k}_n^T &= \textbf{e}_y |k| \cos{\theta_n} + \textbf{e}_x |k| \sin{\theta_n} \\
    \textbf{k}_n^R &= - \textbf{e}_y |k| \cos{\theta_n} + \textbf{e}_x |k| \sin{\theta_n} \\
    \textbf{k}_{\mathrm{inc}} &= \textbf{e}_y |k|
\end{aligned}
\end{equation}
with $|k| = 2 \pi f / c_0$ and $\theta_n$ being the refraction angle.
The transmission and the reflection coefficients then follow as
\begin{equation}
\begin{aligned}
    &T_n = \frac{ a^T_n }{ a_{\textrm{inc}} } \\
    &R_n = \frac{ a^R_n }{ a_{\textrm{inc}} } 
\end{aligned}
\end{equation}
and the absorption coefficient as
\begin{equation}
    \alpha = 1 - \sum_{n} |T_n|^2 - \sum_{n} |R_n|^2 , \, n \in \{ 0, \pm1 \}.
\end{equation}
The transmitted energy distribution follows as (Eq.~\eqref{eq:tau})
\begin{equation*}
    \tau^{\textrm{FEM}}_n = \frac{|T_{n}|^2}{|T_{-1}|^2 + |T_{0}|^2 + |T_{+1}|^2}.
\end{equation*}

The material parameters used for air are
equilibrium density $\rho_0 = \unit[1.2]{kg \cdot m^{-3}}$, 
speed of sound $c = \unit[343]{m \cdot s^{-1}}$, 
dynamic viscosity $\mu = \unit[1.814\times10^{-5}]{Pa \cdot s}$, 
bulk viscosity $\mu_{\mathrm{B}} = \unit[1.0884\times10^{-5}]{Pa \cdot s}$,
thermal conductivity $k = \unit[0.025768]{W \cdot m^{-1} \cdot K^{-1}}$, 
heat capacity at constant pressure $C_p = \unit[1005.4]{J \cdot kg^{-1} \cdot K^{-1}}$, 
and ratio of specific heats $\gamma = 1.4$

\subsection{Sample manufacturing using two-photon lithography}
The fabrication of samples was conducted using two-photon polymerization based additive manufacturing.
Processing was carried out using the Photonic Professional GT2 (Nanoscribe GmbH, Germany) with a build volume of $\unit[10 \times 10 \times 0.8] {cm^3}$ \cite{NanoGuide.23Nov21}.
The voxel shape is elliptic, its size ($\unit[0.1]{\mu m} < r_v < \unit[5]{\mu m}$ and $\unit[0.3]{\mu m} < z_v < \unit[15]{\mu m}$ \cite{NanoGuide.23Nov21}) depends on the illumination process parameters and the optical properties of the photoresist.
The required structures can be created by scanning the voxel through the photoresist following positional data from CAD models. 
Excess photoresist was removed
leaving the cured structures, which were detached from the substrate after drying.


Focusing was conducted by means of immersion objectives $10 \times$ numerical aperture (NA) 0.3 and $25 \times$ NA 0.8 for the grating holder and grating sample, respectively. The photoresists IP-Q and IP-S were used, respectively (Nanoscribe GmbH, Germany).
The development of the photoresist was performed by 1-Methoxy-2-propylacetat (\unit[20]{min}, \unit[60]{ml}) and 2-propanol (\unit[5]{min}, \unit[60]{ml}).
Drying was carried out in air at $\unit[21.7 \pm 0.44]{^{\circ}C}$ and $\unit{32.1 \pm 8.2} {\%}$ relative humidity under a glass hood.
The resulting mechanical properties for IP-S are
Poisson's ratio $\nu_{\textrm{IP-S}} \approx 0.3$,
Young's modulus $E_{\textrm{IP-S}} \approx \unit[5.11]{GPa}$, and
density $\rho_{\textrm{IP-S}} \approx \unit[1.22] {g \, cm^{-3}}$\cite{NanoGuide.23Nov21}.

\subsection{Experimental setup}

As shown in Fig.~\ref{fig:D0_and_exp_setup}d, the measurement setup consists of an assembly that connects a sound source with a microacoustic metagrating. In addition, the assembly is mounted in a position system (not shown) that enables the assembly to move relative to a microphone.
The sound source is an in-house designed and fabricated capacitive micro-machined ultrasound transducer~\cite{Koch.4272021}, which is mounted on TO-18 (transistor-style metal case, see Fig.~\ref{fig:D0_and_exp_setup}d, brassy) and clamped into the frame (see Fig.~\ref{fig:D0_and_exp_setup}d, pink) using the screw (see Fig.~\ref{fig:D0_and_exp_setup}d, green).
In order to measure ultra-high frequency ultrasound, the optical microphone Eta450 Ultra (Xarion, Austria) with a specified frequency range from \SI{50}{\kilo\Hz} to \SI{2}{\mega\Hz} is used~\cite{Fischer.2016, XARIONLaserAcousticsGmbH.2019}.
The frequency range in this work could be extended to \unit[2.2]{MHz} since only relative sound pressure values are used.

To measure radial scans in one plane motorized stages (Physical Instruments, Germany) are used. 
Since the microphone is aligned with the same side to the axis of rotation on the metagrating, the directional characteristic of the microphone are neglected~\cite{Preisser.2016}.
The measured angle lies within $\ang{-70} < \theta < \ang{+70}$ due to prevent collision between the rotatable holder and the microphone attachment.
To reduce the measurement time, the radial axis is constantly moved and triggers sound pressure measurements at the chosen resolution of \ang{0.1}.
In order to accurately adjust the tilt and the positioning of the assembly including the source and the grating relative to the coordinate system of the motorized stage an adjustment laser is used.
The same adjustment laser is used to ensure that the metagrating is in the center of the rotation.
To determine the distance between transducer and the microphone a time of flight measurement with an pulsed stimulation and without metagrating is performed.
The measurements are done with continuous stimulation (sinusoidal signal \unit[10]{Vpp} + DC bias \unit[20]{V}).
The metagrating is aligned with the rotation axis, which is orthogonal to the measurement plane, while center of the microphone is located at the half of the height of the metagrating.
The refracted wave is then measured using the optical microphone on a circular trajectory on a radius of \unit[9.87]{mm} from the metagrating surface.

\subsection{Experimental data analysis}

The raw measured pressure amplitude is fitted using a Gaussian mixture model of the form
\begin{equation}
\begin{aligned}
    &G \left( \theta \right) = \sum_{n} \hat{p}_{n} g_n \left( \theta \right)\\
    &n \in \left\{0, \pm 1 \right\},
\end{aligned}
\end{equation}
with
\begin{equation}
    g_n \left( \theta \right) = e^{-\frac{(\theta - \theta_n)^2}{2 \sigma^2}}
\end{equation}
where $\hat{p}_{n}$ is the pressure amplitude at the beam center and $\sigma$ corresponds to the beam width.
The pressure magnitude error is estimated according to
\begin{equation}
    \Delta \hat{p}_n =
        \sqrt{ \int_{\Theta} \left( p_{exp} \left( \theta \right) - \hat{p}_n g_n (\theta) \right) ^2 w_n \left( \theta \right) \mathrm{d}\Theta }
\end{equation}
with the weighting function $w_n \left( \theta \right) = g_n \left( \theta \right)$.
The pressure amplitude $\hat{p}_{n}$ is then used to determine the normalized transmission within the diffraction orders
\begin{equation}
    \tau^{\textrm{exp}}_n = \frac{\hat{p}_{n}^2}{\sum_m \hat{p}_{m}^2},\,m \in \{ 0, \pm1 \}
\end{equation}
which is linked to Eq.~\eqref{eq:tau}.
This quantity is introduced because of the difficulty of measuring absolute power and taking multiple reflections into account.
Subsequently, the propagation of the error is determined as
\begin{equation}
    \Delta\tau^{\textrm{exp}}_{i} = 2
    \sqrt{\frac{\hat{p}_{i}^4 \left(\Delta\hat{p}_j^2 \hat{p}_j^2 + \Delta\hat{p}_{k}^2 \hat{p}_{k}^2 \right) + \Delta \hat{p}_{i}^2 \hat{p}_{i}^2 \left( \hat{p}_j^2 + \hat{p}_{k}^2 \right)^2
    }{\left(\hat{p}_{i}^2 + \hat{p}_{j}^2 + \hat{p}_{k}^2\right)^4}}
\end{equation}
with $i \in \{ -1, 0, +1 \}$, $j \in \{ 0, +1, -1 \}$, and $k \in \{ +1, -1, 0 \}$,
leading to the errorbars in Figs.~\ref{fig:all_results}beh.

\section*{Acknowledgements}

Y.K.C., S.M., and D.P. acknowledge support from the Universities Australia DAAD joint research co-operation scheme project 57446203 and support from Australian Research Council Discovery Project DP200101708.
All authors thank Gunjankumar Gediya and Tim Brändel for their support in the experimental work.


\clearpage

\end{document}